\documentclass[aps, eqsecnum, preprint]{revtex4}
\usepackage{amsmath}
\usepackage{bm}
\usepackage[dvips]{graphicx}
\usepackage[dvips]{graphics}
\usepackage{subfigure}
\usepackage{epsfig}
\def \d {DNA}
\def \th {\theta}
\def \thi {\theta_{int}}
\def \thx {\theta_{ext}}
\begin{document}
\title{Statistical Mechanics of Thermal Denaturation of DNA oligomers }
\author{Navin Singh and Yashwant Singh}
\affiliation{Department of Physics, Banaras Hindu University, \\ 
Varanasi - 221005, INDIA} 
\email{ysingh@bhu.ac.in}
\begin{abstract}
Double stranded \d\ chain is known to have nontrivial elasticity. We study the
effect of this elasticity on the denaturation profile of \d\ oligomer by
constraining one base pair at one end of the oligomer to remain in unstretched
(or intact) state. The effect of this constraint on the denaturation profile
of the oligomer has been calculated using the Peyrard-Bishop Hamiltonian. The
denaturation profile is found to be very different from the free (i.e. without
the constraint) oligomer. We have also examined how this constraint affects
the denaturation profile of the oligomer having a segment of defect sites
located at different parts of the chain.
\end{abstract} 
\maketitle

\section{Introduction}
DNA is one of the most complex and important biomolecule as it is central
to all living beings. It contains all the informations needed for birth,
development, living and probably sets the average life. Structurally it is a
giant double stranded linear molecule with length ranging from 2$\mu$ m for
simple viruses to $3.5 \times 10^7 \mu m$ for more complex organism
\cite{styr}. How such a molecule came into being during the evolution of life
and how it acquired the ability to store and transmit the genetic informations
are still a mystery. The recent progress in genome mapping and
availability of experimental techniques to study the physical properties of a
single molecule \cite{smith,bock1} has, however, made the field very active
from both biological and physical point of views.

A DNA molecule is not just a static object but a dynamical system having
rather complex nature of internal motions \cite{yak}. The structural elements
such as individual atoms, group of atoms (bases, sugar rings, phosphates)
fragments of double chain including several base pairs, are in constant
movement and this movement plays an important role in the functioning of the
molecule. The solvent in which the molecule is immersed acts as a thermal bath
and provides energy for different motion. In addition, collision with the
molecules of the solution which surrounds DNA, local interactions with
proteins, drugs or with some other ligands also lead to internal motion. These
motions are distinguished by activation energies, amplitudes and
characteristic times. The motions which are of our interest here are opening
of base pairs, formation of bubbles along the chain and unwinding of helix
(denaturation). The energy involved for these motions is of the order of 5-20
kcal/mole. These motions are activated by increasing temperature, increasing
pH of the solvent action of denaturation agents etc. The time scale of these
motions are of the order of microseconds and is therefore generally
unobservable in atomistic simulations as these simulations are restricted to
time scale of nanosecond because of computational cost.

The nature of thermal denaturation leading to separation of two strands has
been investigated for several decades \cite{Wart,Theo2}. Experimentally a
sample containing molecules of a specific length and sequence is prepared.
Then the fraction of bound base pairs as a function of temperature, referred
to as the melting curve, is measured through light absorption, typically at
about 260 nm. For heterogeneous DNA, where the sequence contains both AT and
GC pairs, the melting curve exhibits a multistep behaviour consisting of
plateaus with different  sizes separated by sharp jumps. These jumps have been
attributed to the unwinding of domains characterized by different frequencies
of AT and GC pairs. The sharpness of the jump in long DNA molecules suggests
that the transition from bound to unbound is first-order. The understanding of
this remarkable one-dimensional cooperative phenomena in terms of standard
statistical mechanics, i.e., a Hamiltonian model with temperature independent
parameters is a subject of current interest \cite{Theo1}.

\section{Model Hamiltonian and its Properties}

Since the internal motion that is basically responsible for denaturation is
the stretching of bases  from their equilibrium position along the direction of
the hydrogen bonds that connect the two bases, a \d\ molecule can be
considered as a quasi one dimensional lattice composed of $N$ base pair units.
The forces which stabilize the structure are the hydrogen bonds between
complementary bases on opposite strand and stacking interactions between
nearest neighbour bases on opposite strands. Each base pair is in one of the
two states: either open(non hydrogen bonded) or intact (hydrogen bonded). A
Hamiltonian model that has been found appropriate  to include  these
interactions and describe the displacement of bases leading to denaturation is
the Peyrard-Bishop model (PB model) \cite{Pey1}. The PB model is written as
$$ H = \sum_{i=1}^N H(y_i, y_{i+1}) $$
\begin{equation}
H(y_i, y_{i+1}) = \frac{p_i^2}{2m} + w(y_i, y_{i+1}) + V(y_i)
\end{equation}
where $m$ is the reduced mass of the base pair, $y_i$ denotes the stretching of
the hydrogen bonds connecting the two bases of the $i^{th}$ pair and
$$
p_i = m\left(\frac{dy_i}{dt}\right)
$$
The on-site potential $V(y_i)$ describes the interaction of the two bases of
the $i^{th}$ pair. The Morse potential
\begin{equation}
V(y_i) = D_i(e^{-ay_i} - 1)^2
\end{equation}
which is taken to represent the on-site interaction represents not only the
hydrogen bonds connecting two bases belonging to opposite strands, but also
the repulsive interactions of the phosphates and the surrounding solvent
effects. The flat part at large values of the displacement of this potential
emulates the tendency of the pair "melt" at high temperatures as thermal
phonons drive the molecules outside the well and towards the flat portion of
the potential.

The stacking interactions are contributed by dipole-dipole interactions,
$\pi$-electron systems, London dispersion forces and in water solution, the
hydrophobic interactions. These forces result in a complex interaction pattern
between overlapping base pairs, with minimum energy distance close to 3.4
${\rm \AA}$ in the normal \d\ double helix. The following anharmonic potential
model mimic these features of the stacking energy:
\begin{equation}
W(y_i, y_{i+1}) = \frac{1}{2} k [ 1 + \rho\exp(-\alpha(y_i+y_{i+1}))]
(y_i-y_{i+1})^2 
\end{equation}
where $k$ is the coupling constant and the second term in the bracket
represents the anharmonic term. When due to stretching the hydrogen bonds
connecting the bases break, the electronic distribution on the bases is
modified causing the stacking interactions to decrease. This is
taken into account by the exponential term in Eq.(2.3). It may be noted that
the effective coupling constant decreases from $k(1+\rho)$ to $k$ when either
one of the two interacting base pairs is stretched. This decrease in coupling
provides a large entropy in the denaturation. The parameter $\alpha$ in
Eq.(2.3) defines the "anharmonic range". 

\begin{figure*}[h]
\includegraphics[width=3.5in]{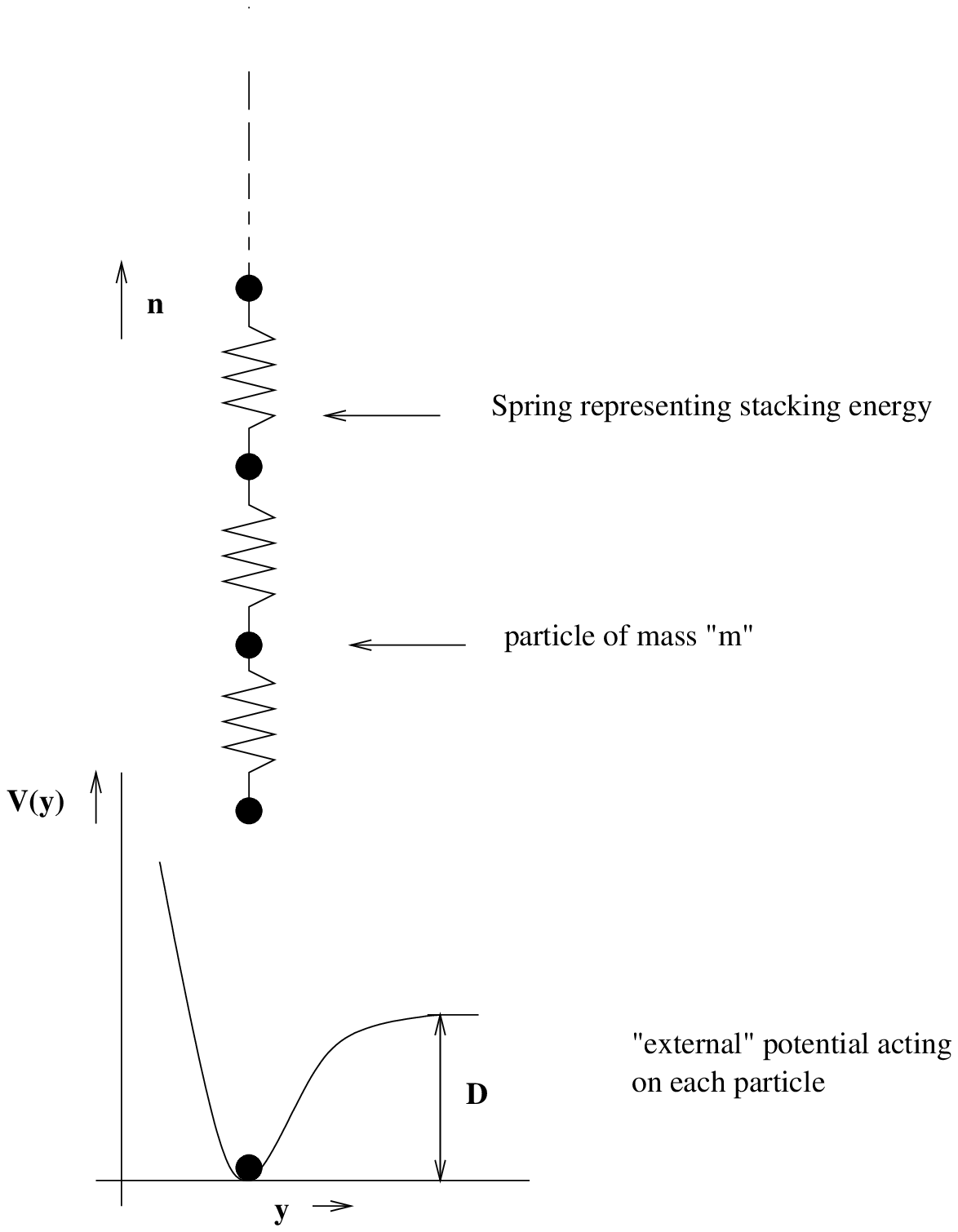}
\caption\small{Schematic of quasi-one dimensional model of DNA}
\end{figure*} 

The model described above can be viewed as a model of one-dimensional
monoatomic lattice (see Fig. 1) with each atom having mass $m$ and nearest
neighbour interaction given by Eq.(2.3). Furthermore, each atom is subjected to
"external" potential given by Eq.(2.1) which effect is to confine the chain
in the potential well. The melting takes place because of competition between
the thermal energy which leads to displacement and "external" field and the
nearest neighbour interactions which lead to the confinement. The model,
therefore, represents a one-dimensional system that differs from the usual
one-dimensional systems that do not show phase transitions.

The model Hamiltonian of Eq.(2.1) has extensively been used to study the
melting profile of a very long ($N\to\infty$) homogeneous \d\ chain using both
statistical mechanical calculations and constrained temperature molecular
dynamics \cite{Pey2, Pey3}. Analytical investigation of nonlinear
dynamics of the model suggests that intrinsic energy localization can initiate
the denaturation \cite{Pey4}. The model for a long homogeneous chain exhibits a
peculiar type of first-order transition with finite melting entropy, a
discontinuity in the fraction of bound pairs and divergent correlation
lengths. However, as the value of the stacking parameter $\alpha$ increases
and the range of the "entropy barrier" becomes shorter than or comparable to
the range of the Morse potential the transition changes to second order. The
crossover is seen at $\alpha/a = 0.5$ \cite{Theo1}. Though the PB model seems
capable of explaining the multistep melting in a sequence-specific disorder
\cite{cule}, how this disorder affect the nature of transition has yet to be
understood. In other work the PB model has been used to understand the melting
profile of short chains \cite{campa} and effect of defects on this profile
\cite{nav}. 

\section{Denaturation Profile}

In a given system of \d\ chains the average fraction $\theta$ of bonded base
pairs can be written as $\th = \thx\thi$. $\thx$ represents the average
fraction of the strands forming duplexes (double strands), while $\thi$ is the
average fraction of unbroken bonds in the duplexes. The equilibrium
dissociation of the duplexes $C_2$ to single strand $C_1$ may be represented
by the relation $C_2 \rightleftharpoons 2 C_1$. The dissociation equilibrium
can be neglected in the case of long chains as $\thx$ is practically 1 while
$\thi$ and therefore $\th$ goes to zero. This is because in the case of long
\d\ chains when $\th$ goes practically from 1 to zero near the denaturation
transition, while most bonds are disrupted and the \d\ has denatured, few
bonds still remaining  prevent the two strands getting apart from each
other. It is only at $T >> T_m$ ($T_m$ being the melting temperature at which
half of the bonds are broken) there will be real separation. Therefore at the
transition the double strand is always a single molecule and in calculation
based on PB model one has to calculate only $\thi(\equiv \th)$. On the
contrary, in the case of short chains the processes of single bond disruption
and strand dissociation tend to happen in the same temperature range,
therefore, the computation of $\thx$ in addition to $\thi$ is essential. 

Unfortunately, at present, we do not have any reliable method for calculating
$\thx$. The method which has been used is based on the partition function of
rigid molecules and adjustable parameters \cite{campa, Wart} to be determined
from experimental data. To avoid this shortcoming of the theory we in the
present article discuss the denaturation profile of oligonucleotides of given
sequence with a base pair at one end of the chain held in such a way that it
remains  at its equilibrium seperation(i.e. there is no stretching) at  all
temperatures. This can be done by creating a deep potential well for this
base pair or attaching the one end of both strand to a substrate. This
will be referred to as chain with constraint in order to distinguish it form
the "free chain". One of the advantage of having a constraint of this type
is obvious; the problem of divergence of the partition function for the PB
model for short chains no more exists. 

The \d\ molecule is known to have non trivial elastic properties: When the two
strands of \d\ molecule are pulled apart by applying a force at one end of the
chain, a novel phase transition is found (in case of infinitely long chain)
to take place at which the two strands are pulled completely apart
\cite{nelson}. The phase digram plotted in the plane of temperature and force
reveals the elastic properties of the \d\ chain. Our study reported in this
paper differs from the situation just described in two ways; (i) a short chain
of 21 given base sequence is considered and (ii) instead of pulling  the chain
apart the end base pair is constrained to be in unstreched or intact position. 

The oligonucleotide which we consider has the following sequence given by: 
\begin{eqnarray}
 ^{5'}A C G C T A T A C T C A C G T T A A C A G_{3'}  \nonumber \\
 _{3'}T G C G A T A T G A G T G C A A T T G T C^{5'}
\end{eqnarray}
The denaturation profile of this oligonucleotide has been studied by Campa and
Giansanti \cite{campa} and by us \cite{nav}. We take the same parameters as
in the previous study. Thus, $D_{AT} = 0.05$ eV, $D_{GC} = 0.075$ eV, $a_{AT}
= 4.2 {\rm \AA^{-1}}$, \; $a_{GC} = 6.9{\rm \AA^{-1}}$, $k = 0.025 \;{\rm eV
\AA^{-2}},\;\rho = 2 \;{\rm and}\; \alpha = 0.35 {\rm \AA^{-1}}$. When one end
of chain is held fixed at $y=0$ distance, the fraction of intact bonds is
found using the relation
\begin{equation}
\th = \frac{1}{N}\sum_{i=1}^N \langle \vartheta(y_0 - y_i)\rangle
\end{equation}
where $\vartheta(y)$ is Heaviside step function, $N$ the total number of
base pairs and the canonical average $\langle\cdot\rangle$ is evaluated by
considering all configuration of the chain with one end fixed. The $i^{th}$
bond is considered bound if the value of $y_i$ is smaller than a chosen
threshold $y_0$. For $y_0$ we have taken a value of 2 ${\rm \AA}$. Since the
model Hamiltonian of Eq.(2.1) couples only the nearest neighbours the
calculation of canonical average reduces to multiplication of finite matrices.
The discretization of the coordinate variables and introduction of a proper
cut-off on the matrices and the number of base pairs in the chain the number
of matrices to be multiplied.

\begin{figure*}[h]
\includegraphics[width=5.5in]{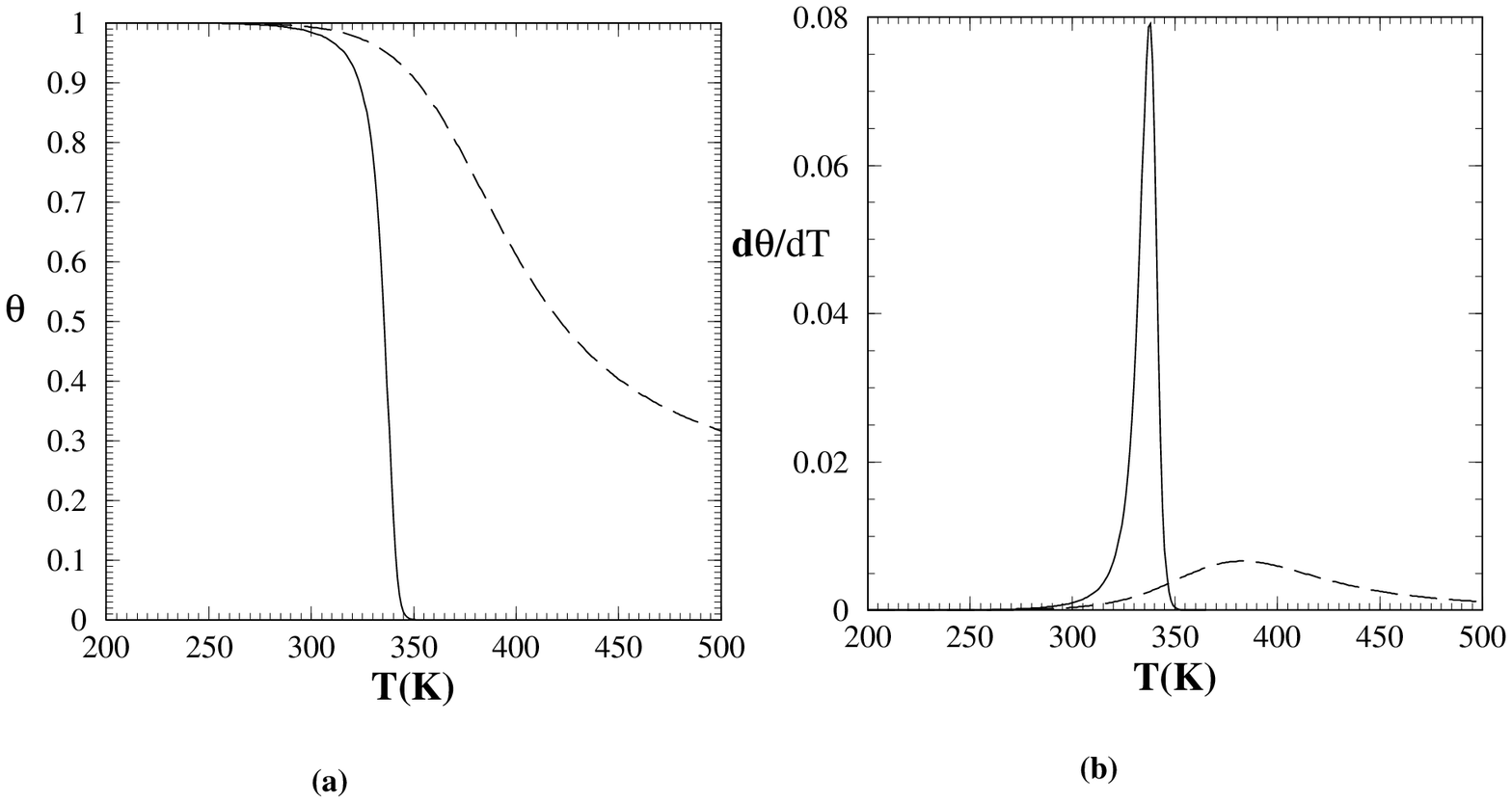}
\caption\small{Variation of $\theta$ and $d\th/dT$ as a function of temperature
for free (solid line) and constrained chain (dashed line)}
\end{figure*} 

In Fig.2 we compare the value of $\th$ as a function of temperature for the
free and the constrained chain. The influence of the constraint on the
denaturation profile is enormous. While in case of free chain the breaking of
bonds takes place in a narrow temperature range, in the constrained chain it
is over a wide temperature range. The value of temperature at which $d\th/dT$
is maximum shifts about 45$^0$C and the peak is much wider and smaller
compared to that of free chain.

Next we calculated the denaturation profile for the following two nucleotides
having a segment of chain with defect sites:
\begin{eqnarray}
(a) \; && ^{5'}A C G C T A T A C T C A C G T T A A C A G _{3'}  \nonumber \\
       && _{3'}T C G C T A T A C T C T G C A A T T G T C ^{5'}   \\
(b) \; && ^{5'}A C G C T A T A C T C A C G T T A A C A G _{3'} \nonumber \\
       && _{3'}T G C G A T T A C T C A C G T T T T G T C ^{5'}
\end{eqnarray}

\begin{figure*}[h]
\includegraphics[width=5.5in]{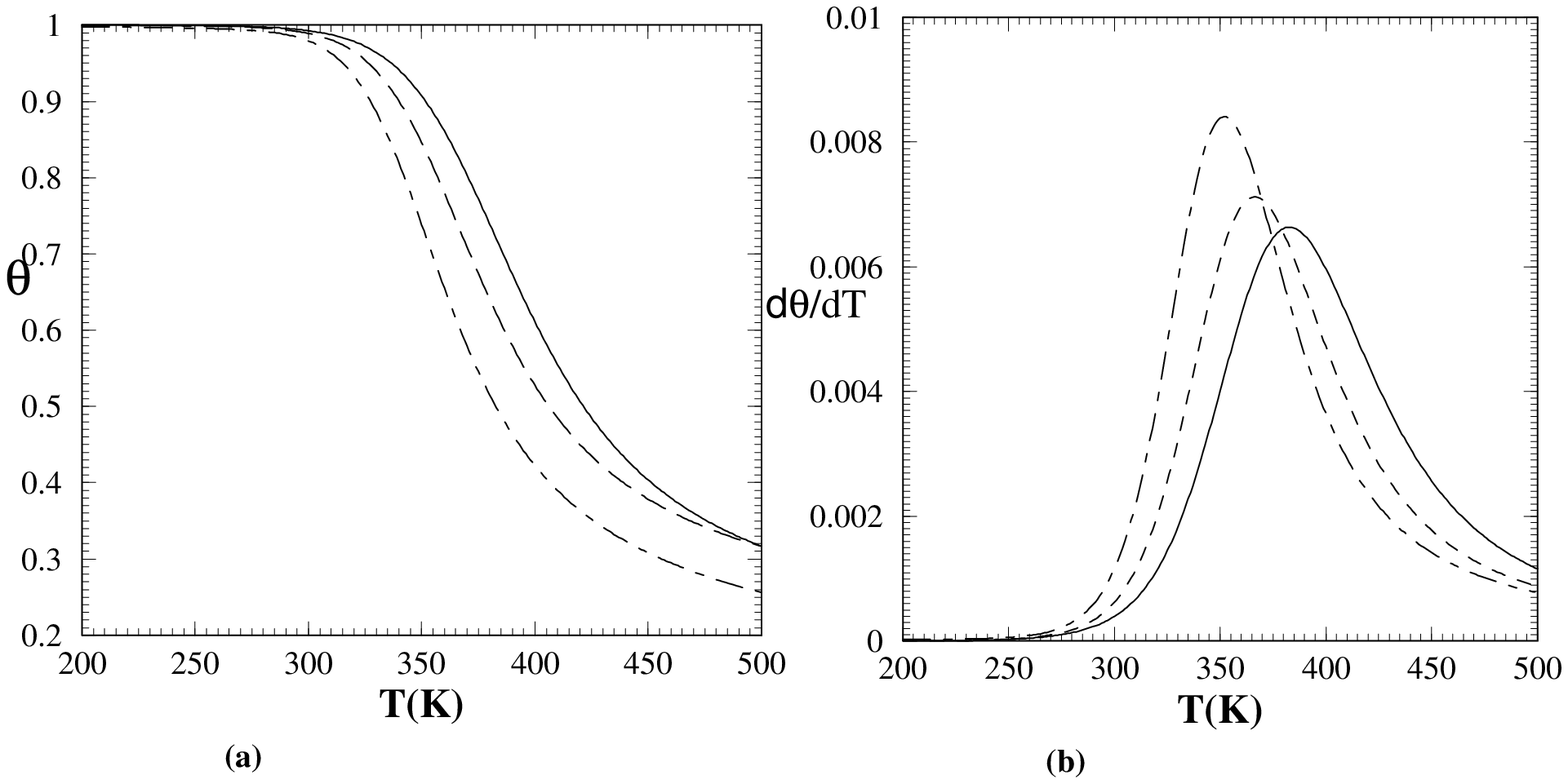}
\caption\small{Fig.(a): Variation of $\th$ as a function of temperature for the
constrained chain without defect (solid line) with 10 defect at one end of the
chain (dashed line) and 10 defects in the middle (dot-dashed line). Fig.(b):
Plot of $d\th/dT$ for the same.}
\end{figure*} 

While both oligonucleotide have ten defect sites their locations differ. In
(a) the defects are on the left end from site 2 to 11 while in (b) it is in
middle from site 6 to 16. The position of the base pair is counted from left.
The purpose is to see how these defects affect the denaturation profile and
the formation of loop and stem  as often seen in a single strand \d\ or RNA.
In Fig.3 we plot the variation of $\th$ and $d\th/dT$ as a function of
temperature. While the denaturation takes place at low temperature the shift
is more when the defects are in the middle.

\begin{figure*}[h]
\includegraphics[width=5.5in]{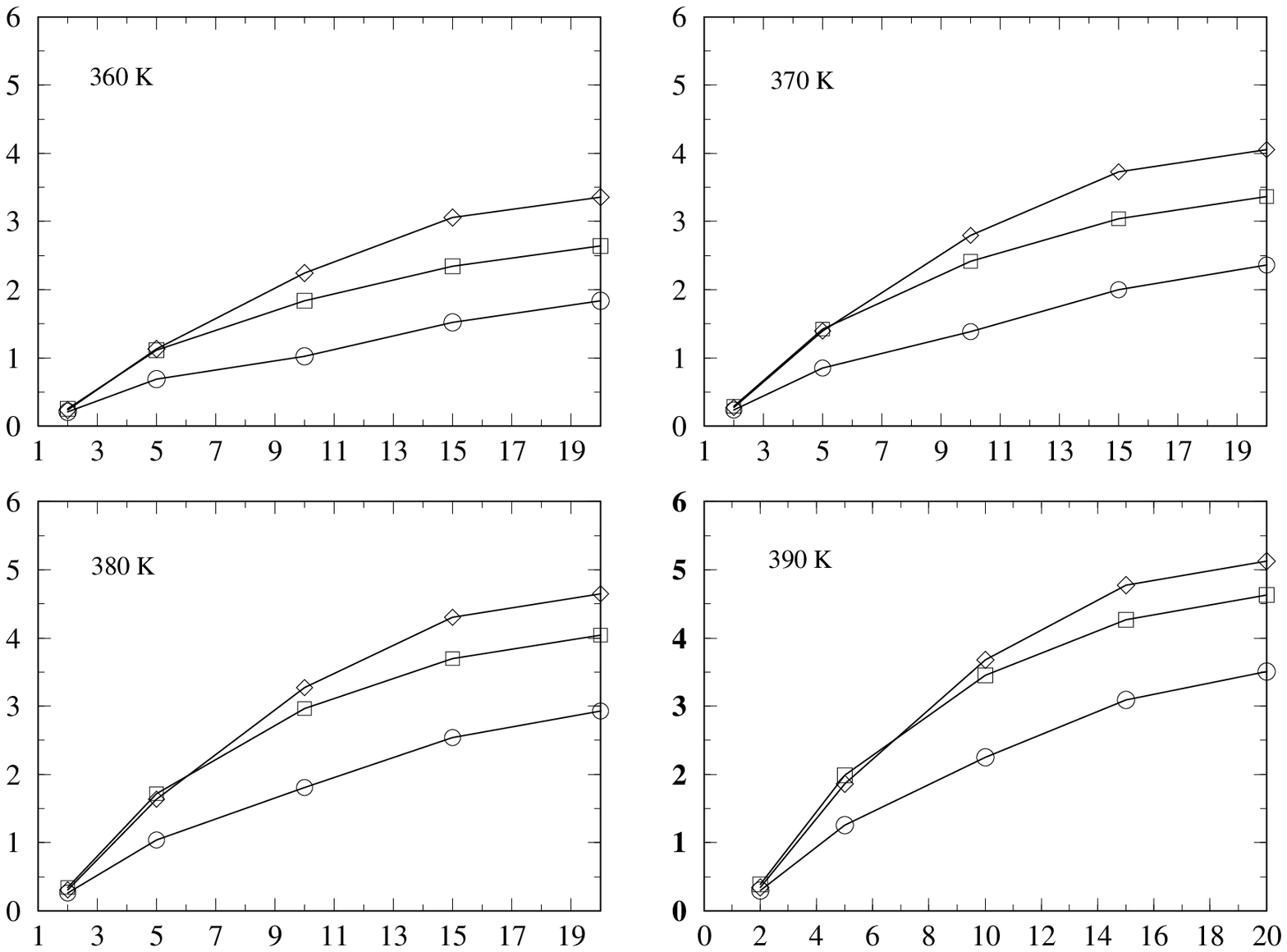}
\caption\small{Plot of $\langle y_n \rangle$ vs. $n$ (site position) at four
different temperatures. {\it Circles} denotes the constrained chain without
defect, {\it square} denotes the chain with 10 defect at one end while {\it
diamond} denotes the chain with 10 defect in the middle.} 
\end{figure*}

The other quantities of interest are the mean value of displacement of
$n^{th}$ base pairs defined as 
\begin{equation}
\langle y_n \rangle = \frac{1}{Z} \int \left( \prod_{i=1}^{N} dy_i \right)
y_n \exp [-\beta\sum_{i=1}^N H(y_i, y_{i+1})]
\end{equation}
and its fluctuations
\begin{equation}
\langle |\delta y_n|^2 \rangle = \frac{1}{Z} \int \left( \prod_{i=1}^{N} dy_i
\right) \left( y_n - \langle y_n \rangle \right)^2\exp [-\beta\sum_{i=1}^N
H(y_i, y_{i+1})] 
\end{equation}
Here $Z = \int \left( \prod_{i=1}^{N} dy_i \right)\exp [-\beta\sum_{i=1}^N
H(y_i, y_{i+1})] $ is the partition function of the chain. 

\begin{figure*}[h]
\includegraphics[width=2.5in]{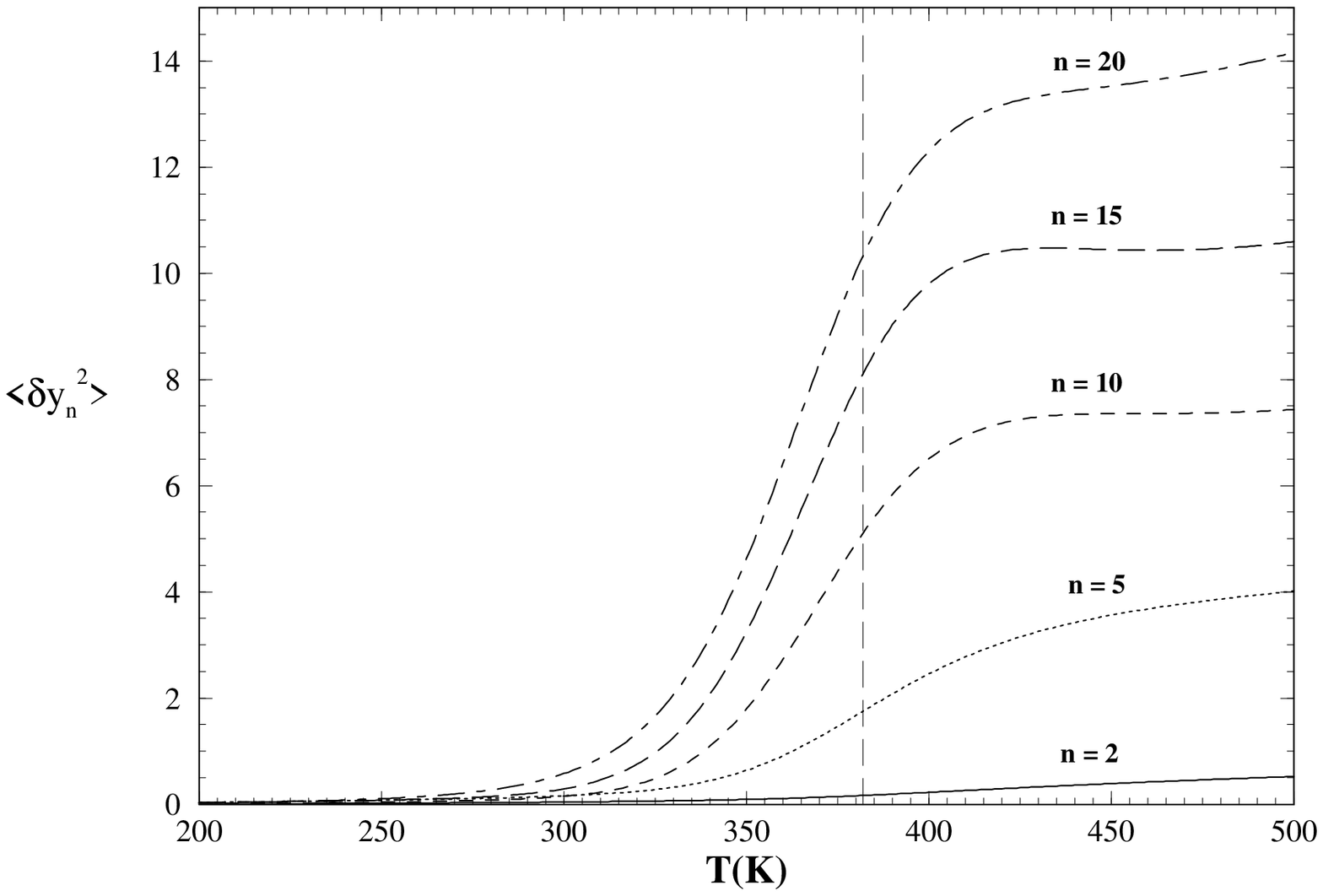}
\caption\small{Plot showing the variation of transverse correlation length 
with temperature for different value of $n$.} 
\end{figure*} 

Because of the constraint that the first base pair of the chain is in
unstreched condition the  value of $\langle y_n \rangle $ as well as $
\langle |\delta y_n|^2 \rangle$ depend on the site $n$. In Fig. 4 we
plot the value of $\langle y_n \rangle $ as a function of $n$ at several
temperatures for all three constrained chain as described above. We find
in all cases the opening of the chain starts from the open end of the chain.
In Fig. 5 we plot $\langle |\delta y_n|^2 \rangle$  as a function of
temperature for several base pairs. The quantity $\langle |\delta y_n|^2
\rangle$ measures the transverse correlation length for the base pair $n$.
This correlation length remains almost zero for all $n$ when the
oligonucleotide is in the native state but at the denaturation its value 
increases. At a given temperature the value of $\langle |\delta y_n|^2
\rangle$ depends on $n$ and it increases with $n$. For a long chain we
expect it to diverge for large value of $n$. 

In conclusion we wish to emphasize that the effect of constraining a base pair
at one end of a given oligonucleotide has very significant effect on the
denaturation profile. 

The work has been supported through research grant by Department of Science
and Technology and Council of Scientific and Industrial Research, New Delhi,
Government of India.

\end{document}